\documentclass [12pt]{article}
\newcommand{\mypar}{\bigskip \noindent}
\begin{document}

\begin{center}
\begin{Large}
{\bf  From Wormholes to the Warp Drive:\\ Using theoretical
physics to place ultimate bounds on technology\\}
\end{Large}
\end{center}
\vspace*{0.50cm}
\begin{center}
\vspace*{1.00cm}\begin{large} {William A. Hiscock\\} {
Department of Physics\\
Montana State University\\
Bozeman, Montana 59717\\}
\end{large}
[e-mail: hiscock@physics.montana.edu]\\
\end{center}

\bigskip
\bigskip
\begin{abstract}
The serious study of such science fiction staples as wormholes,
time travel, and the warp drive, as a means of understanding and
constraining possible realistic solutions within General
Relativity is reviewed.
\end{abstract}
\newpage

\bigskip\noindent

\begin{center}
\begin{large}
{\it "You cannot change the laws of physics, Captain." }\\
-- Lt. Commander Montgomery Scott, Chief Engineer\\ U.S.S.
Enterprise  NCC-1701
\end{large}
\end{center}
\vspace*{2.00cm}

\noindent Wormholes. Time travel. The ``warp drive''. These
staples of science fiction have today become subjects of serious
study by theoretical physicists. Each of these concepts can be
identified with mathematical solutions to the equations of
Einstein's theory of general relativity - the modern theory of
gravity, in which the phenomena of gravity are explained in terms
of the geometry of curved spacetime. By studying these solutions
in detail, theoretical physicists seek to better learn what bounds
exist on the behavior of matter and geometry in our present
physical theories. Such studies can yield new insights into the
developing quantum theory of gravity, and can also help establish
ultimate limits on technology - not based on the degree of scope
or difficulty of an engineering challenge, but by compatibility
with our core understanding of the fundamental laws of physics.

\mypar Three decades ago, adequate theories were first developed
for three of the four fundamental forces of nature: the so-called
``strong'' and ``weak'' nuclear forces which operate on subatomic
scales, and the electromagnetic force which is responsible for
most of what we experience in everyday life: chemistry, biology,
and technology (non-nuclear). The theoretical model of these three
forces has been so successful in matching experiment that it has
been termed the ``Standard Model'' of particle physics. Professor
Howard Georgi of Harvard has testified to the robustness of the
model in saying ``Everything anyone has ever seen can be explained
in terms of the Standard Model.'' While there are some highly
interesting aspects of the Standard Model where new developments
in particle physics are still taking place (e.g., the study of
neutrino properties), the development of the Standard Model of
particle physics must rank as one of the premier scientific
accomplishments of the twentieth century.

\mypar The fourth fundamental force in nature, the gravitational
force, remains in many ways an enigma. Gravity was the first force
for which a detailed mathematical theory was developed, by Isaac
Newton, over three hundred years ago. That theory is still
adequate to successfully navigate our planetary spacecraft, such
as Voyagers 1 and 2, around the solar system. Einstein developed
our modern theory of gravity not because Newton's theory had been
found to disagree with experiment, but because Newton's theory was
mathematically incompatible with the dictates of Einstein's
special theory of relativity, which requires that all observers
measure the same value for the speed of light, c, regardless of
their state of motion (this feature of nature has been
overwhelmingly confirmed by experiment and observation).
Einstein's new theory, finalized in 1916, is usually called
``general relativity'', a name that unfortunately obscures the
fact that it is actually a theory of the gravitational
interaction.

\mypar Einstein's theory is widely regarded as one of the most
beautiful theories in all of science. It describes gravity in
terms of the curved geometry of the four dimensional (three space,
one time) world we inhabit. Physicist John Wheeler has beautifully
summarized the content of general relativity in a sentence:
``Matter tells space how to curve; space tells matter how to
move''. The first part of the sentence tells us the origin of
spacetime curvature. Matter (in any form, including energy, as per
Einstein's famous $E = mc^2$) is the source of gravity, causing
the curvature of space (or ``spacetime'', a shorthand for our
four-dimensional world). The second part of the sentence describes
how matter behaves in this curved four dimensional world, namely
that it follows paths which are the curved space generalization of
``straight lines'' - geodesics. The ``great circle'' route of an
airplane flying from New York to Tokyo is an example of a
geodesic; it is not a straight line (which would pass through the
solid Earth), but is the shortest path connecting the two cities
along the curved surface of the Earth.

\mypar Despite these great strides that have been made in the
understanding of the behavior of matter and energy at the
fundamental level, there are still many unresolved questions to
attract the attention of physicists. Perhaps the greatest of these
is the search for an understanding of how gravity fits together
with the other three fundamental forces. It is clear that a truly
fundamental theory of gravity must be quantum in nature, and
general relativity is not. The contrast between the Standard
Model, built on the foundation of quantum field theory, and
general relativity, highly accurately tested but still inadequate
because of its classical nature, has caused many to term the
theory of quantum gravity to be the ``holy grail'' of theoretical
physics. However, despite decades of effort, little is year truly
clear about the nature of quantum gravity. Should gravity be able
to stand alone in a quantum theory, independent of the other
fields that constitute nature, or can gravity only be understood
in terms of a unified theory that attempts to handle all aspects
of matter and its interactions at once? M-theory (sometimes called
``the theory formerly known as superstrings'') is a bold attempt
at a theory of everything, solving the problem of the quantum
theory of gravity by merging all matter and interactions into a
single theory in which the fundamental objects are not point
particles, but are higher dimensional objects such as loops of
``superstring''.

\mypar
In ordinary matter, quantum effects become quite evident when one
reaches the atomic scale, at roughly $10^{-10}$ meters. Today's
high energy particle accelerators allow us to study the behavior
of matter and energy at scales as small as $10^{-17}$ meters,
where the existence of particles such as quarks, gluons, and W and
Z bosons are evident that are completely invisible back up at the
atomic scale. Quantum gravitational effects are expected to become
important when one reaches the Planck scale, about $10^{-35}$
meters. If superstring theory describes nature, this is the size
of a typical loop of superstring material. The large difference in
scale between the Planck scale of superstrings and today's
experiments - some 18 orders of magnitude, a billion billion - is
one  reason why it is so difficult for physicists working on
superstring models to make experimental predictions from their
theory that can be tested with our present level of technology.

\mypar
The even larger span between the scales at which quantum effects
become evident in matter, and the Planck scale of quantum gravity,
twenty-five orders of magnitude, is a realm where the
semiclassical theory of gravity may be a good tool to help us
understand how gravity and quantum physics are to be married. In
semiclassical gravity, the gravitational field is still treated
classically, being described by the curved spacetime of general
relativity. The matter which creates the spacetime curvature,
however, is treated using quantum field theory. The resulting
theory cannot be ultimately acceptable, since it mixes classical
and quantum physics, but it should be a valuable tool and
approximate description of the behavior of nature over length
scales (and energy scales) that reach down towards the realm where
string theory may provide an ultimate exegesis.

\mypar Theoretical physicists expand their understanding of the
strengths, limits, and behavior of a particular theory by
``pushing'' the theory to its limiting cases. For example, we
understand general relativity much better by studying the
properties of black hole solutions to the Einstein equations than
we would if we restricted attention to the tiny differences
between the predictions of Newtonian gravity and general
relativity for, say planetary orbits. These tiny differences
provide valuable experimental tests of GR, since we don't have any
black holes in our near neighborhood that we can study in detail;
but understanding the full range of behavior allowed by the theory
is best enhanced by examining the nature of its predictions for
situations that are far from ``normal'' - e.g., for black holes,
and much more speculative solutions to the equations such as those
called ``wormholes'', the ``warp drive'', and those which lead to
the possibility of time travel.

\mypar Each of these concepts can be identified with particular
spacetime geometries, and hence with mathematical solutions to
Einstein's theory of general relativity. In fact, any
four-dimensional geometry for a spacetime can be termed a solution
of Einstein's theory - the question physics must answer is whether
the matter necessary to curve spacetime into a particular
``shape'' is of a form compatible with nature as we understand it.
If one postulates a spacetime geometry with some exotic property
(e.g., a wormhole), then it is likely that ``exotic'' matter is
necessary to generate that geometry. By ``exotic matter'' we mean
matter that has properties not usually seen in ordinary situations
- such as negative mass or energy densities. Such properties are
so far from those exhibited by normal matter that we must
seriously ask whether such behavior is compatible with the quantum
field theory description of matter in the Standard Model. This
then provides a link between extreme quantum behavior of matter
and exotic geometries, and hence insight into the possibilities of
quantum gravity. Next, one may ask whether the exotic geometry is
stable - would small changes destroy the exotic structure, for
example by closing off a wormhole. In some cases, the small change
necessary to drive the instability is provided by quantum
mechanics, again providing a link between these exotic spacetime
structures and quantum aspects of gravity. A pencil balanced on
its point is an example of a mathematical solution of the laws of
mechanics which is unstable; the smallest motion will cause it to
fall in one direction or another.  The goal is to continue to make
the model of the geometry more realistic, adding more aspects of
physics to its description, and determining whether the original
exotic geometry is compatible with this more realistic treatment.
If physics rules the feature out, then no amount of clever
engineering can hope to turn science fiction to fact; on the other
hand, if no incompatibility with known physics is found, then it
might be possible for future engineers to create such geometric
structures in spacetime. It is important to note that the work of
physicists does not aim to place limits on the potential scope of
engineering, except where violations of well-tested and accepted
laws of physics is involved. Simply because an engineering
solution might involve energies and scales many orders of
magnitude beyond what Earthly technology can achieve today does
not daunt the inquiring physicist at all.

\mypar An example of how such a study proceeds is provided by the
story of how the present interest in these concepts largely began
- surprisingly enough, with science fiction, namely the writing of
the novel ``Contact'' by Carl Sagan. Sagan asked Kip Thorne, the
Feynman Professor of Physics at Caltech, for advice to help ensure
that the method chosen to transport the novel's heroine across the
Galaxy would not be scientifically ludicrous. Thorne suggested
replacing the notion of diving through a black hole as a portal to
distant realms with the idea of using a wormhole.

\mypar A black hole is a spherical region of spacetime surrounded
by an ``event horizon'', which functions as a one-way ``gate''.
Anything that crosses the event horizon into the black hole -
rocks, spaceships, light itself - cannot escape the black hole. In
the simplest model of a black hole, where the black hole has no
spin or electric charge, any matter falling into the hole is
inevitably crushed into a spacetime singularity, where the
gravitational tidal forces are infinitely strong. There is no
escape. The notion that black holes might provide ``tunnels to
elsewhere'' first arose in the 1960's, when physicists found that
if a black hole had an electric charge, or if it was spinning,
matter (e.g., a spaceship) falling into the hole could avoid being
crushed into a singularity and would eventually emerge from the
confines of the hole into ``another universe''. This ``other
universe'' might be identified with a distant region of our own
universe, which led quickly to science fiction authors using black
holes as interstellar shortcuts through spacetime. However, the
physics did not stop there: through the 1970's, work by a large
number of physicists (including Nobel Laureate Subrahmanyan
Chandrasekhar) showed that the ``tunnel'' structure inside such
models of black holes is unstable, both classically and quantum
mechanically. By including more physics in the model, we find that
the tunnel is closed off, replaced by a singularity as in the
simpler black hole case. Thus, by studying these exotic models of
black hole interiors in greater detail, the ``science fiction''
aspects were actually ruled out when nature was examined more
closely.

\mypar For this reason, Thorne suggested that Sagan use a wormhole
in his novel, rather than a black hole. A ``wormhole'' is, as the
name suggests, a topologically distinct route between two
locations which are the ``mouths'' of the wormhole (think of an
actual wormhole which has two ends on the surface of an apple). A
priori, the distance through the wormhole, from mouth to mouth,
could be either longer or shorter than the conventional distance
traveled through ordinary space (or over the surface of the
apple). For the purposes of science fiction, wormholes that
provide shortcuts through spacetime are of course preferred.
Unlike a black hole, wormholes do not generally possess event
horizons, there are no ``one-way gates'' in spacetime involved.

\mypar The serious study of wormhole geometries in general
relativity began in the late 1980's when Thorne and his students,
and subsequently many other theoretical physicists, showed that
wormholes inevitably must involve ``exotic matter''. In order to
hold the ``throat'' of the wormhole (between the two mouths) open,
there must be some form of matter or energy present which has
negative mass or energy. Every form of classical matter known to
physics has positive mass, hence the term "exotic" for this
alternate possibility. This is where the truly interesting
research questions begin, for when one combines a quantum
description of matter and energy with curved spacetime, it is
known that situations involving negative mass can arise. This has
even been demonstrated in the laboratory, in the Casimir effect, a
very weak attractive force that exists between two uncharged
parallel plates made of electrically conducting material.  This
force exists due to the negative energy in the volume between the
two plates, which is caused by the conducting plates eliminating
some of the usual vacuum modes between them. While this is an
example of how quantum effects can create an effective negative
mass, it is a very weak effect - the negative mass is small in
magnitude (much smaller than the  mass of the plates) and occupies
a very small volume of space. For wormholes, the key question
still being investigated today is whether any plausible
configuration of quantum fields in curved spacetime could create
sufficient negative mass to hold open a macroscopic wormhole - one
large enough for a starship, a person, or even a single atom to
pass through. So far, the evidence suggests that it is highly
unlikely such structures could exist: strongly negative mass such
as that required to hold a wormhole open has never been observed
and is considered unlikely to exist (even antimatter has positive
mass).  The final word here is certainly not yet written; our
understanding continues to improve as new models incorporating
more particle physics are examined.

\mypar
Physicists are also interested in microscopic wormholes, much
smaller than a proton, down at the scale of the Planck length.
These might play an important role in quantum gravity, where
spacetime is often viewed as a foaming sea of tiny wormholes,
forming and disappearing every instant. Another significant
problem for anyone enthusiastic about wormholes is the "worm"
problem: how do you create a wormhole? Perhaps a small, quantum
wormhole (if they exist) could be caught and expanded to useful
size. This important question has not really been addressed yet in
any satisfactory manner.

\mypar
The "warp drive" is another example of an interesting geometry
that is yielding new insights to general relativity. In 1994,
Miquel Alcubierre, at the time a graduate student at the
University of Wales in Cardiff,  published a mathematical
description of a spacetime geometry that embodies the properties
usually associated in science fiction with a "warp drive". In this
geometry, a "starship" can apparently travel faster than the speed
of light, traversing interstellar distances of many lightyears in
an arbitrarily short time - both as measured by those on the
starship, and those at the destination. I say "apparently",
because the starship never exceeds the speed of light as measured
by a local observer - the basic tenet of Einstein's special
relativity is not violated. The motion of the ship occurs because
the spacetime in front of the starship contracts while that behind
the starship expands, transporting the starship forward. Further
study has shown that such solutions are probably all unphysical,
for several reasons. The foremost difficulty was noted by
Alcubierre in his original paper: such solutions inevitably
involve a large concentration of negative mass, which appears to
be just as implausible in this context as in the case of
wormholes. Still, research continues in this area, as there are
enough significant variations on Alcubierre's original idea to
keep theoreticians busy for some time to come.

\mypar
Is time travel possible? Ignoring the pedestrian everyday
progression of time, the question can be divided into two parts:
Is it possible, within a short time (less than a human life span),
to travel into the distant future? And is it possible to travel
into the past? Our current understanding of fundamental physics
tells us that the answer to the first question is a definite yes,
and to the second, maybe.

\mypar
The mechanism for traveling into the distant future is to use the
time-dilation effect of special relativity. Special relativity
teaches us that time is not absolute and universal for all
possible observers: a moving clock will appear to tick more slowly
the closer it approaches the speed of light. This effect, which
has been overwhelmingly supported by experimental tests, applies
to all types of clocks, including biological aging. Departing from
Earth in a spaceship that could accelerate continuously at a
comfortable one {\it g} (an acceleration that would produce a
force equal to the gravity we feel everyday at the earth's
surface), one would begin to approach the speed of light relative
to the Earth within about a year. As the ship continued to
accelerate, it would come ever closer to the speed of light, and
its clocks would appear to run at an ever slower rate relative to
the Earth. Under such circumstances, a round trip to the center of
our Galaxy and back to the Earth-- a distance of some 60,000
light-years--could be completed in only a little more than 40
years of ship time. Upon arriving back at the Earth, the astronaut
would be only 40 years older, while 60,000 years would have passed
on the Earth. (Note that there is no ``twin paradox,'' because it
is unambiguous that the space traveler has felt the constant
acceleration for 40 years, while a hypothetical twin left behind
on an identical spaceship circling the Earth has not.) Such a trip
would pose formidable engineering problems: the amount of fuel
required, even assuming a perfect 100\% efficient conversion of
mass into energy, greatly exceeds the mass of a planet. But
nothing in the known laws of physics would prevent such a trip
from occurring.

\mypar
Time travel into the past is a much more uncertain proposition.
There are many solutions to Einstein's equations of general
relativity, including some involving wormholes in motion, that
allow a person to follow a timeline that would result in her
encountering herself--or her grandmother--at an earlier time. The
problem, as we've seen before, is deciding whether these solutions
represent situations that could occur in the real universe, or
whether they are mere mathematical oddities incompatible with
known physics. Much work has been done by theoretical physicists
in the past decade to try to determine whether, in a universe that
is initially without time travel, one can build a time machine--in
other words, if it is possible to manipulate matter and the
geometry of space-time in such a way as to create new paths that
circle back in time. The main possible roadblock to creating a
time machine seems to be a quantum instability of the spacetime -
any attempt to "turn on" a time machine would result in spacetime
changing its geometry to prevent the machine's operation. This
instability was first discovered by Deborah Konkowski of the U.S.
Naval Academy and myself in 1982. Once again, the question of the
possibility of time travel remains open and an active area of
research; there are serious questions as to whether this quantum
instability is truly strong enough to prevent time travel, and
whether there are not special quantum states that avoid the
instability altogether.

\mypar
These concepts associated with science fiction continue to be
useful areas of research for real, serious, theoretical physics.
By studying these solutions to Einstein's equations, we can gain
new insights about the ultimate limits of gravity and its relation
to quantum field theory descriptions of matter.

\vspace*{3.00cm}

\noindent
{\bf Suggestions for additional reading:}

\mypar
{\it Black Holes \& Time Warps Einstein's Outrageous Legacy}, K.
S. Thorne (Norton, N.Y., 1994).

\mypar
{\it Dreams of a final theory}, S.Weinberg (Pantheon, N.Y., 1993).

\mypar
{\it Time Travel in Einstein's Universe: The Physical
Possibilities of Travel Through Time}, J. Richard Gott (Houghton
Mifflin, N.Y., 2001).

\mypar
{\it Lorentzian Wormholes: From Einstein to Hawking}, M. Visser
(Springer-Verlag 1996).

\end{document}